\documentclass[conference]{IEEEtran}

\IEEEoverridecommandlockouts
\usepackage{amsfonts}
\usepackage{amsmath}
\usepackage{algorithmic}
\usepackage[boxed]{algorithm}
\usepackage{mathrsfs}
\usepackage{graphicx}

\begin{document}
\title{Robust Linear Processing for Downlink Multiuser MIMO System With Imperfectly Known Channel}

\author{\IEEEauthorblockN{Pengfei Ma, Xiaochuan Zhao, Mugen Peng, Wenbo Wang }
\IEEEauthorblockA{Beijing University of Posts and Telecommunications \\
Beijing, China}}

\maketitle

\begin{abstract}
%\boldmath
In practical systems, due to the time-varying radio channel, the
channel state information (CSI) may not be known well at both
transmitters and receivers. For most of the current multiuser
multiple-input multiple-output (MIMO) schemes, they suffer a
significant degression on the performance due to the mismatch
between the true and estimated CSI. To alleviate the performance
penalty, a robust downlink multiuser MIMO scheme is proposed in this
paper by exploiting the channel mean and antenna correlation. These
channel statistics are more stable than the imperfect CSI estimation
in the time-varying radio channel, and they are used, in the
proposed scheme, to minimize the total mean squared error under the
sum power constraint. Simulation results demonstrate that the
proposed scheme effectively mitigates the performance loss due to
the CSI mismatch.
\end{abstract}


\IEEEpeerreviewmaketitle

\section{Introduction}
The multiple-input multiple-output (MIMO) system,
employing multiple transmit and receive antennas, has been
recognized as an effective way to improve the spectral efficiency of
the radio channel \cite{1} \cite{2}. More recently, multiuser
schemes have been investigated for MIMO systems to further improve
the multiuser sum capacity.

Early studies have assumed a perfectly knowledge of the channel
state information (CSI) available at the transmitter. \cite{3}
extended the single-user scheme \cite{4} to the multiuser system.
However, without exploring the multiuser channel information, it
simply treated the multiuser interference as the white noise. The
scheme in \cite{5}, on the contrary, utilized the multiuser
information effectively to minimize the total mean squared error
(TMMSE) and, naturally, possessed a better performance.

The CSI can be obtained at the transmitter either by using a
feedback channel from the receiver to the transmitter in frequency
division duplex (FDD) systems, or by invoking the channel
reciprocity in time division duplex (TDD) systems. However, using
feedback in FDD systems, the limited resources for the feedback,
associated with the propagation delay and schedule lag, heavily
degrade the accuracy of the CSI at the transmitter. As to the
channel reciprocity in TDD systems, the accuracy of the CSI is
corrupted by antenna calibration errors and turn-around time delay.
In respect that the performance would degrade significantly under
the imperfect CSI, it is necessary to design a multiuser scheme
which is stable to the imperfect CSI.

In robust design methodologies, Maxmin (worst-case) and Bayesian
(stochastic) are two well known ones \cite{7}. The former optimizes
the performance under the worst case of random channels, thus, it is
so conservative that its average performance is even worse than
non-robust schemes \cite{8}. The latter maximizes the ensemble
average performance over a pre-described stochastic distribution of
the CSI. When the stochastic distribution matches well with the true
CSI, the latter outperforms the former.

The scheme in \cite{7} was a Bayesian design for downlink multiuser
MIMO systems with the imperfectly known CSI. It introduced a channel
error matrix to the cost function of \cite{3}, then found the
solution which minimized the average cost. However, similar with
\cite{3}, the multiuser interference was also treated as the white
noise. Therefore, it is expected that the performance can be
improved by exploring the multiuser information.

In this paper, a robust scheme for downlink multiuser MIMO systems
is proposed based on the TMMSE criterion. A more general channel
model involving the channel mean and antenna correlation is
considered. The scheme is a Bayesian design which minimizes the
average cost function under the sum power constraint.

The rest of this paper is organized as follows. The channel model
and problem formulation are described in Section II. The Section III
presents the design of the robust multiuser scheme for the
correlated imperfect known channel under the sum power constraint.
Simulation results and analysis are given in Section IV. Finally,
the Section V concludes the paper.

\emph{Notation}: Boldface upper-case letters denote matrices, and
boldface lower-case letters denote column vectors. $tr( \cdot )$, $(
\cdot )^*$, $( \cdot )^H$, $|| \cdot ||_2$ and $|| \cdot ||_F$
denote trace, conjugate, conjugate transposition, Euclidian norm and
Frobenius norm, respectively. $E( \cdot )$ represents the
expectation of a stochastic process. $[{\bf{ \cdot }}]_{i,j}$,
$[{\bf{ \cdot }}]_{{\bf{:}},j}$ denote the $(i$,$j)$-th element and
$j$-th column of a matrix, respectively.


\section{Problem statement}
\subsection{Channel Model}
Consider a base station (BS) with $M$ antennas and $K$ mobile
stations (MS's) each having $N_i (i = 1 \dots K)$ antennas.
Represented by a matrix ${\bf{H}}_i \in \mathcal{C}^{N_i \times M}$,
the downlink MIMO channel to MS$_i$ is assumed to be frequency-flat
and quasi-static block fading. Suppose a non-zero-mean channel with
both transmit and receive antenna correlations, ${\bf{H}}_i$ is
written as follows \cite{9}\cite{11}
\begin{equation}
{\bf{H}}_i  = \sqrt {\frac{{W_i }}{{W_i  + 1}}} {\bf{\tilde
H}}_{0,i}  + \sqrt {\frac{1}{{W_i  + 1}}} {\bf{R}}_{0,r,i}
^{\frac{1}{2}} {\bf{\Delta }}_i {\bf{R}}_t ^{\frac{1}{2}}
\end{equation}
where $W_i$ is the ratio of the power in the mean component to the
average power in the variant component of ${\bf{H}}_i$; ${\bf{\Delta
}}_i \in \mathcal{C}^{N_i \times N_i}$ is random, we assume that its
entries form an independent identical distribution (i.i.d.) complex
Gaussian collection with zero-mean and identity covariance, i.e.,
${\bf{\Delta }}_i \sim \mathcal{CN}(0,1)$; ${\bf{\tilde H}}_{0,i}
\in \mathcal{C}^{N_i \times M}$ is the normalized channel mean, and
${\bf{R}}_{0,r,i} \in \mathcal{C}^{N_i \times N_i}$ and ${\bf{R}}_t
\in \mathcal{C}^{M \times M}$ are the normalized correlation
matrices of the receiver of MS$_i$ and the transmitter of BS,
respectively. (1) is rewritten into the following for simplicity
\cite{11}
\begin{equation}
{\bf{H}}_i  = {\bf{\tilde H}}_i  + {\bf{R}}_{r,i} ^{\frac{1}{2}}
{\bf{\Delta }}_i {\bf{R}}_t ^{\frac{1}{2}}
\end{equation}
where ${\bf{\tilde H}}_i  = \sqrt {W_i /(W_i  + 1)} {\bf{\tilde
H}}_{0,i}$ is the channel mean, and ${\bf{R}}_{r,i}  = 1/(W_i  +
1){\bf{R}}_{0,r,i}$ is the equivalent correlation matrix of the
receiver of MS$_i$.

The channel mean and correlation are more stable than the
instantaneous channel information, and they are usually acquired by
time-averaging on channel measurements. In the Rayleigh channel, for
example, the non-zero channel mean ${\bf{\tilde H}}_i$ is obtained
by averaging channel measurements over a window of tens of the
channel coherence time \cite{10}. Furthermore, the channel model (2)
can also denote the correlated Rician MIMO channel, in which case
the channel mean represents the line-of-sight (LOS) component of the
MIMO channel.

In this paper, we assume that transmitters and receivers only know
channel means and antenna correlations.

\subsection{Problem Formulation}
We assume that there are $L_i (i = 1 \dots K)$ substreams between BS
and MS$_i (i = 1 \dots K)$, that is to say, BS transmits $L_i$
symbols to MS$_i$ simultaneously. Then the signal received at MS$_i$
is
\begin{equation}
{\bf{y}}_i  = {\bf{A}}_i ^H {\bf{H}}_i \sum\limits_{k = 1}^K
{{\bf{B}}_k {\bf{x}}_k }  + {\bf{A}}_i ^H {\bf{n}}_i
\end{equation}
where ${\bf{y}}_i  \in \mathcal{C}^{N_i \times 1}$
  is the received signal vector, and ${\bf{x}}_i  \in \mathcal{C}^{L_i \times 1}$
  is the transmitted signal vector from BS to MS$_i$ with zero-mean and normalized covariance matrix $\mathbf{I}$.
We assume the transmitted signal vectors of different users are
uncorrelated, i.e., $E\left( {{\bf{x}}_i {\bf{x}}_j^H } \right) =
\delta_{ij}\bf{I}$, where ${\delta_{ij}}$ is the Kronecker function,
${\delta_{ij} = 1}$, when ${i=j}$ and ${\delta_{ij}= 0}$, when
${i\neq j}$. We also assume the noise vector is independent of any
signal vector. A linear post-filter ${\bf{A}}_i \in \mathcal{C}^{N_i
\times L_i} (i = 1 \dots K) $ is used at MS$_i$ to recover an
estimation of the transmitted signal vector ${\bf{x}}_i$.
${\bf{H}}_i$ defined in (1) [or (2)] denotes the MIMO channel from
BS to MS$_i$. ${\bf{B}}_i \in \mathcal{C}^{M \times L_i} (i = 1
\dots K) $ is used at BS to weight the transmitted signal vector
${\bf{x}}_i$. After passing through ${\bf{B}}_i$, ${\bf{x}}_i$
becomes into an $M \times 1$ signal vector which is transmitted by
$M$ transmit antennas of BS. ${\bf{n}}_i  \in \mathcal{C}^{N_i
\times 1}$ is the noise vector with the correlation matrix
${\bf{R}}_{n_i} = \sigma _n ^2  {\bf{I}}_{{\bf{N}}_i}$, where
${\bf{I}}_{{\bf{N}}_i}$ denotes the ${{\bf{N}}_i} \times
{{\bf{N}}_i}$ identity matrix. In this paper, we assume $L_1= \dots
=L_K=L$.

$\mathbf{A}_i$ and $\mathbf{B}_i$ $(k=1 \dots K)$ are jointly
designed to minimize the total MSE under the sum power constraint.
Hence, we get

\begin{equation}
\begin{array}{cl}
   {\min } & {TMSE = E\left( \sum\limits_{k = 1}^K {||{\bf{x}}_k  - {\bf{y}}_k ||^2 } \right) }  \\
   {s.t.} & {tr\left(\sum\limits_{k = 1}^K {{\bf{B}}_k {\bf{B}}_k ^H } \right) \le P}  \\
 \end{array}
\end{equation}
where $P$ is the total transmit power of BS.

\section{Robust TMMSE Scheme}
According to (3), the $j$-th user's MSE is
\begin{equation}
\begin{array}{lll}
 MSE_j  & = & E\left( {||{\bf{x}}_j  - {\bf{y}}_j ||^2 } \right)   \\
 & = & E(tr({\bf{A}}_j ^H {\bf{H}}_j (\sum\limits_{i = 1}^K {{\bf{B}}_i {\bf{B}}_i ^H }){\bf{H}}_j ^H {\bf{A}}_j  + \sigma _n ^2 {\bf{A}}_j ^H {\bf{A}}_j \\
 & & - {\bf{B}}_j ^H {\bf{H}}_j ^H {\bf{A}}_j - {\bf{A}}_j ^H {\bf{H}}_j {\bf{B}}_j  + {\bf{I}}))\\
\end{array}
\end{equation}
Substitute (2) into (5) and note $E\left( {\bf{H}}_i \right)=
{\bf{\tilde H}}_i $, hence $E\left( {\bf{R}}_{r,i} ^{\frac{1}{2}}
{\bf{\Delta }}_i {\bf{R}}_t ^{\frac{1}{2}} \right)= {\bf{0}}$, we
obtain
\begin{equation}
\begin{array}{lll}
 MSE_j & = & tr({\bf{A}}_j ^H {\bf{\tilde H}}_j (\sum\limits_{i = 1}^K {{\bf{B}}_i {\bf{B}}_i ^H } ){\bf{\tilde H}}_j ^H {\bf{A}}_j  + \sigma _n ^2   {\bf{A}}_j ^H {\bf{A}}_j
 \\& {} & - {\bf{B}}_j ^H {\bf{\tilde H}}_j ^H  {\bf{A}}_j -  {\bf{A}}_j^H{\bf{\tilde H}}_j {\bf{B}}_j  + {\bf{I}}) + E(tr({\bf{A}}_j ^H
\\& {} &{\bf{R}}_{r,j}^{\frac{1}{2}} {\bf{\Delta}}_j{\bf{R}}_{t}^{\frac{1}{2}}(\sum\limits_{i = 1}^K {{\bf{B}}_i {\bf{B}}_i ^H } )
{\bf{R}}_{t} ^{\frac{1}{2}H} {\bf{\Delta }}_{j}^H {\bf{R}}_{r,j} ^{\frac{1}{2}H} {\bf{A}}_j )) \\
 \end{array}
\end{equation}
Observe the last part in (6)
\begin{equation}
\begin{array}{ll}
 E(tr({\bf{A}}_j ^H {\bf{R}}_{r,j}^{\frac{1}{2}} {\bf{\Delta}}_j{\bf{R}}_{t}^{\frac{1}{2}}(\sum\limits_{i = 1}^K {{\bf{B}}_i {\bf{B}}_i ^H } )
{\bf{R}}_{t} ^{\frac{1}{2}H} {\bf{\Delta }}_{j}^H {\bf{R}}_{r,j} ^{\frac{1}{2}H} {\bf{A}}_j ))  \\
 {\;\;\;} = {\;} tr({\bf{R}}_{t}^{\frac{1}{2}} (\sum\limits_{i = 1}^K {{\bf{B}}_i {\bf{B}}_i ^H } ){\bf{R}}_{t}^{\frac{1}{2}H}
 E({\bf{\Delta }}_{j}^{H} {\bf{R}}_{r,j}^{\frac{1}{2}H} {\bf{A}}_j {\bf{A}}_j^H {\bf{R}}_{r,j}^{\frac{1}{2}} {\bf{\Delta }}_{j})) \\
 \end{array}
\end{equation}

Moreover, as ${\bf{\Delta }}_i \sim \mathcal{CN}(0,1)$ with i.i.d.
entries, $E(\left[{\bf{\Delta }}_j\right]_{:,n} (\left[{\bf{\Delta
}}_j\right]_{:,m})^H ) = \delta _{n,m} \bf{I}$. Therefore, the
$(m$,$n)$-th entry of the expectation in the right side of (7) is
\begin{equation}
\begin{array}{llll}
 E(\left[ {\bf{\Delta }}_{j}^{H} {\bf{R}}_{r,j}^{\frac{1}{2}H} {\bf{A}}_j {\bf{A}}_j^H {\bf{R}}_{r,j}^{\frac{1}{2}} {\bf{\Delta }}_{j} \right]_{m,n}) & {} &\\
 {\;\;\;\;\;\;\;\;\;\;\;\;\;}
 \begin{array}{llll}
 = & E((\left[ {\bf{\Delta }}_j \right]_{:,m})^H {\bf{R}}_{r,j}^{\frac{1}{2}H} {\bf{A}}_j {\bf{A}}_j^H {\bf{R}}_{r,j}^{\frac{1}{2}} \left[ {\bf{\Delta }}_j \right]_{:,n} ) \\
 = & tr({\bf{R}}_{r,j}^{\frac{1}{2}H} {\bf{A}}_j {\bf{A}}_{j}^H {\bf{R}}_{r,j}^{\frac{1}{2}} E(\left[{\bf{\Delta }}_j\right]_{:,n} (\left[{\bf{\Delta }}_j\right]_{:,m})^H )) \\
 = & \delta_{n,m}tr( {\bf{R}}_{r,j}^{\frac{1}{2}H} {\bf{A}}_j {\bf{A}}_j ^H {\bf{R}}_{r,j}^{\frac{1}{2}} ) \\
 = & \delta_{n,m}tr( {\bf{A}}_j^H {\bf{R}}_{r,j} {{\bf{A}}_j} ) \\
\end{array}
\end{array}
\end{equation}
Thus, the expectation in the right side of (7) is
\begin{equation}
 E({{\bf{\Delta }}_{j}^H {\bf{R}}_{r,j}^{\frac{1}{2}H} {\bf{A}}_j {\bf{A}}_j^H  {\bf{R}}_{r,j}^{\frac{1}{2}}
  {\bf{\Delta }}_j}) =  tr( {\bf{A}}_j^H {\bf{R}}_{r,j} {{\bf{A}}_j} ) {\bf{I}}
\end{equation}
 Substitute (9) into (7), we obtain
\begin{equation}
\begin{array}{lll}
 E(tr({\bf{A}}_j ^H {\bf{R}}_{r,j}^{\frac{1}{2}} {\bf{\Delta}}_j{\bf{R}}_{t}^{\frac{1}{2}}(\sum\limits_{i = 1}^K {{\bf{B}}_i {\bf{B}}_i ^H } )
{\bf{R}}_{t} ^{\frac{1}{2}H} {\bf{\Delta }}_{j}^H {\bf{R}}_{r,j} ^{\frac{1}{2}H} {\bf{A}}_j ))  \\
{\;\;\;} = {\;} tr({\bf{R}}_{t}^{\frac{1}{2}} (\sum\limits_{i = 1}^K
{{\bf{B}}_i {\bf{B}}_i ^H } ) {\bf{R}}_{t}^{\frac{1}{2}H}
 tr( {\bf{A}}_j^H {\bf{R}}_{r,j} {{\bf{A}}_j} ) {\bf{I}} )\\
{\;\;\;} = {\;} tr( {\bf{A}}_j^H {\bf{R}}_{r,j} {{\bf{A}}_j})
tr((\sum\limits_{i = 1}^K {{\bf{B}}_i {\bf{B}}_i ^H } ) {\bf{R}}_t) \\
\end{array}
\end{equation}
Substitute (10) into (6)
\begin{equation}
\begin{array}{lll}
 MSE_j & = & tr({\bf{A}}_j ^H {\bf{\tilde H}}_j (\sum\limits_{i = 1}^K {{\bf{B}}_i {\bf{B}}_i ^H } )
{\bf{\tilde H}}_j ^H {\bf{A}}_j  + \sigma _n ^2 {\bf{A}}_j ^H
{\bf{A}}_j \\
& {} & - {\bf{B}}_j ^H {\bf{\tilde H}}_j ^H {\bf{A}}_j -
{\bf{A}}_j^H {\bf{\tilde H}}_j {\bf{B}}_j  + {\bf{I}}) \\
& {} & + tr({\bf{A}}_j^H {\bf{R}}_{r,j} {{\bf{A}}_j})
tr((\sum\limits_{i = 1}^K {{\bf{B}}_i {\bf{B}}_i ^H } ) {\bf{R}}_t) \\
 \end{array}
\end{equation}
The Lagrangian of (4) is
\begin{equation}
\begin{array}{lll}
L({\bf{A}}_1 , \dots, {\bf{A}}_K ,{\bf{B}}_1 ,\dots, {\bf{B}}_K ) & {} &\\
{\;\;\;\;\;\;\;\;\;\;\;\;\;} =  \sum\limits_{k = 1}^K MSE_k +
\lambda ( {tr(\sum\limits_{k = 1}^K {{\bf{B}}_k {\bf{B}}_k ^H } ) -
P} )
 \end{array}
\end{equation}
where $\lambda$  is the Lagrangian multiplier associated with the
total power constraint. So the Karush-Kuhn-Tucker (KKT) conditions
\cite{10} of (4) are
\begin{eqnarray}
\frac{{\partial L({\bf{A}}_1 ,...,{\bf{A}}_K ,{\bf{B}}_1
,...,{\bf{B}}_K )}}{{\partial {\bf{A}}_i^* }}& = & \bf{0}
\\
\frac{{\partial L({\bf{A}}_1 ,...,{\bf{A}}_K ,{\bf{B}}_1
,...,{\bf{B}}_K )}}{{\partial {\bf{B}}_i ^* }}& = & \bf{0}
\\
\lambda \left({tr(\sum\limits_{i = 1}^K {{\bf{B}}_i {\bf{B}}_i ^H }
) - P} \right)& = & 0
\\
\lambda  & \ge & 0
\end{eqnarray}
Among them, (13)(14) come from the fact that the gradients of the
Lagrangian (12) definitely vanish at the optimal point, and (15) is
known as the complementary slackness. According to (11)$\sim$(14),
we obtain
\begin{equation}
\begin{array}{lll}
{\bf{A}}_i = & ({\bf{\tilde H}}_i (\sum\limits_{k = 1}^K {{\bf{B}}_k
{\bf{B}}_k ^H } ){\bf{\tilde H}}_i ^H  + tr((\sum\limits_{k = 1}^K
{{\bf{B}}_k {\bf{B}}_k ^H } ){\bf{R}}_t)
{\bf{R}}_{r,i}\\
& + \sigma _n ^2 {\bf{I}})^{ - 1} {\bf{\tilde H}}_i {\bf{B}}_i
\end{array}
\end{equation}
\begin{equation}
\begin{array}{lll}
{\bf{B}}_i = & (\sum\limits_{k = 1}^K {{\bf{\tilde H}}_k ^H
{\bf{A}}_k {\bf{A}}_k ^H {\bf{\tilde H}}_k }  + tr(\sum\limits_{k =
1}^K
{{\bf{A}}_k {\bf{A}}_k^H {\bf{R}}_{r,k}}){\bf{R}}_t \\
& +  \lambda {\bf{I}})^{ - 1} {\bf{\tilde H}}_i ^H {\bf{A}}_i
\end{array}
\end{equation}

Substitute (18) into (15) , we can find $\lambda$ is the root of the
equation
\begin{equation}
\lambda (tr\left( {{\bf{X}}({\bf{X}} + {\bf{Y}} + \lambda {\bf{I}}}
\right)^{ - 2} ) - P) = 0
\end{equation}
where
\begin{equation}
{\bf{X}} = \sum\limits_{k = 1}^K {{\bf{\tilde H}}_k ^H {\bf{A}}_k
{\bf{A}}_k ^H {\bf{\tilde H}}_k }
\end{equation}
\begin{equation}
{\bf{Y}} = tr(\sum\limits_{k = 1}^K {{\bf{A}}_k {\bf{A}}_k^H
{\bf{R}}_{r,k}}){\bf{R}}_t
\end{equation}
As ${\bf{R}}_t$ is the normalized correlation matrix of the
transmitter, it is Hermitian, hence both $\bf{X}$ and $\bf{Y}$ are
Hermitian, and so is $\bf{X}+\bf{Y}$. Perform the eigenvalue
decomposition
\begin{equation}
{\bf{X}} + {\bf{Y}} = {\bf{UDU}}^H
\end{equation}
where $\bf{U}$ is unitary and $\bf{D}$ is diagonal. If $\lambda  \ne
0$, (19) can be rewritten to
\begin{equation}
\sum\limits_{n = 1}^M {\frac{{\left[ {{\bf{U}}^H {\bf{XU}}}
\right]_{n,n} }}{{(d_n  + \lambda )^2 }}}  - P = 0
\end{equation}
where $d_n$ is the $n$-th diagonal element of $\mathbf{D}$. Using a
binary search, the root of (23) can be found quickly. Since the
left-hand side of (23) is monotonous in $\lambda$ when $\lambda \ge
0$, the upper and lower bounds on $\lambda$ can be acquired by
replacing $d_n$ with $d_{min}$ and $d_{max}$, respectively. Thus,

\begin{equation}
\lambda _{upper}  = \left( {\sqrt {\frac{{tr(\bf{X})}}{P}} - d_{\min
} } \right)^+
\end{equation}
\begin{equation}
\lambda _{lower}  = \left( {\sqrt {\frac{{tr(\bf{X})}}{P}} - d_{\max
} } \right)^+
\end{equation}
where $(\cdot)^+$ means that the expression takes the value inside
the parentheses if the value is positive, otherwise it takes zero. A
numerical binary search, then, can be carried out between these two
bounds to find the root of (23) up to a desired precision. Once
there is no root between the bounds, which implies that the
inequality constraint (16) is inactive, $\lambda  = 0$ is the only
available solution to (18). From (17) (18), it can be found that the
optimal transmit matrices ${\bf{B}}_k (k = 1\dots K)$ are functions
of the receive matrices ${\bf{A}}_k (k = 1\dots K)$, and vice versa.
Therefore an iterative algorithm to calculate ${\bf{A}}_k$ and
${\bf{B}}_k (k = 1\dots K)$ is proposed as follows.

\begin{algorithm}
\begin{algorithmic}

\STATE{\tt\small Initialize ${\bf{B}}_k^{(0)}$ and
${\bf{A}}_k^{(0)}$ $(k = 1\dots K)$ randomly. \vspace{3pt} \\
$n=0$}\vspace{3pt}

\STATE{\tt 1) {\tt\small {\parindent 6mm Calculate $\lambda$ from
${\bf{A}}_k ^{(n)}$ $(k = 1\dots K)$ by \vspace{1pt}

solving (19).}}\vspace{3pt}

\STATE{\tt 2) {\tt\small {\parindent 6mm  Calculate ${\bf{B}}_k^{(n
+ 1)} (k = 1\dots K)$ from ${\bf{A}}_k ^{(n)} $ \vspace{1pt}

$(k =1\dots K)$ and $\lambda$ using (18).}}} \vspace{3pt}

\STATE{\tt 3) {\tt\small {\parindent 6mm  Calculate ${\bf{A}}_k
^{(n+1)} (k =1\dots K)$ from ${\bf{B}}_k^{(n + 1)}$ \vspace{1pt}

$ (k = 1\dots K)$ using (17).}}} \vspace{3pt} }

\STATE{\tt 4) {\tt\small {\parindent 6mm  Repeat 1), 2) and 3) until
\vspace{1pt}

\parindent 8mm $\sum\limits_{k = 1}^K ({||{\bf{A}}_k ^{(n + 1)} - {\bf{A}}_k^{(n)}||_F^2
 + ||{\bf{B}}_k ^{(n + 1)} - {\bf{B}}_k ^{(n)}||_F^2}) < \varepsilon$.
\vspace{1pt}

\parindent 6mm In our simulation, we set $\varepsilon = 0.0001$.}} \vspace{3pt} }

\end{algorithmic}
\end{algorithm}

\section{Simulation results}
In this section, numerical simulations have been carried out to
evaluate the performance of the proposed scheme.  We assume that the
BS equipped with four antennas ($M=4$) is communicating with two
MS's ($K=2$) each with $N$ receive antennas ($N_1=N_2=N$). Also we
assume that the number of substreams of each MS is equal to $2$
$(L_1=L_2=2)$, moreover, both the two MS's have the same $W_i$
$(W_1=W_2=W)$. QPSK is employed in the simulations and no channel
coding is considered. Let the transmit antenna correlation matrix
${\bf{R}}_t$ be $\left[ {{\bf{R}}_t } \right]_{i,j} = 0.9^{|i - j|}$
and the receive antennas be uncorrelated, i.e., ${\bf{R}}_{r,i} =
{\bf{I}}_{N}$.

\begin{figure}
\centering
\includegraphics[width=3in, height=2.5in]{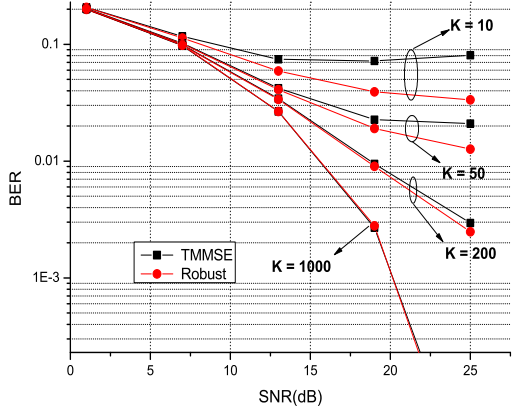}
\caption{Comparison of the BER performance of the robust scheme and
the TMMSE, when $N=2$ and $W=10,50,200,1000$.} \label{Fig 1}
\end{figure}

Firstly, we compare the bit error rate (BER) of the proposed robust
TMMSE scheme with that of the traditional TMMSE scheme. Defining the
signal-to-noise ratio (SNR) as the ratio of total transmitted power
to the noise power of each antenna ($SNR = P/\sigma _n ^2$), Fig.
\ref{Fig 1} is the average BER curves versus the SNR When $N = 2$.
In order to highlight its impact on the BER performance, different
values of $W$ are used in the evaluations. When $W$ is small, the
channel mean poorly reflects the instantaneous channel state, thus
the receiver can not completely eliminate the interference among the
transmitted signals, which further induces an irreducible error
floor at high SNR region. However, the proposed robust scheme
overcomes the traditional one with a noticeable gain. As the $W$
increases, the transmitter obtains more precise CSI, therefore the
residual interference is mitigated greatly and the error floor
vanishes. In addition, the gain between the proposed robust scheme
and the traditional one turns small when the uncertainty of the
channel state is decreasing.

\begin{figure}
\centering
\includegraphics[width=3in, height=2.5in]{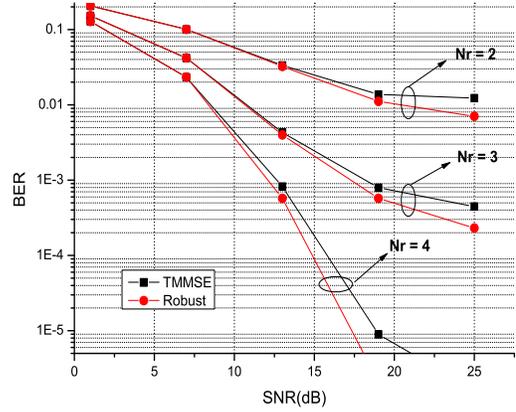}
\caption{Comparison of the BER performance of the robust scheme and
the TMMSE, when $N=2,3,4$ and $W=50$.} \label{Fig 2}
\end{figure}

In Fig. \ref{Fig 2}, we compare the BER performance when the number
of receive antennas is increasing. The additional receive antennas
provide more spatial diversity gain. In this figure, $W$ is fixed to
be $50$ and $N$ changes from $2$ to $4$. Although both the two
schemes explore the additional receive diversity gain, the proposed
robust scheme obviously has a better performance for all $N$'s due
to its insensitivity to the imperfect CSI.

\begin{figure}
\centering
\includegraphics[width=3in, height=2.5in]{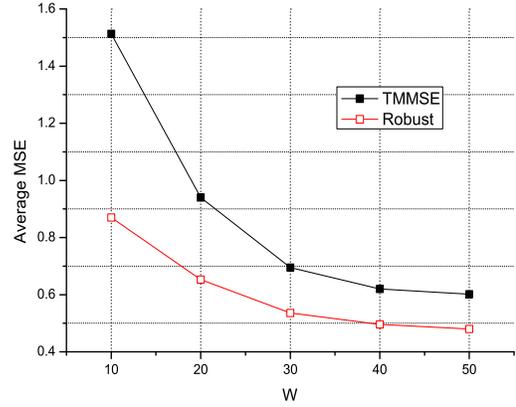}
\caption{Comparison of the average MSE of the robust scheme and the
TMMSE, when $N=2$ and $SNR = 20$dB.} \label{Fig 3}
\end{figure}

\begin{figure}
\centering
\includegraphics[width=3in, height=2.3in]{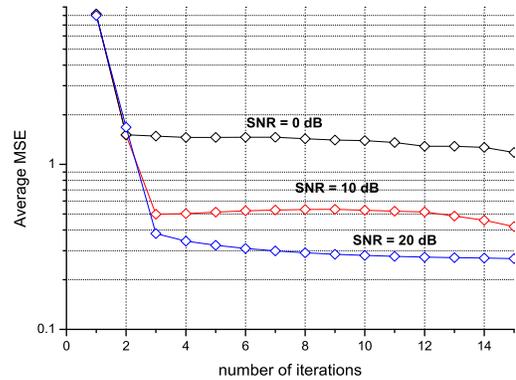}
\caption{Convergence of the robust scheme, when $N=2$ and $W=100$.}
\label{Fig 4}
\end{figure}

Fig. \ref{Fig 3} shows the average MSE as a function of $W$, when
SNR = $20$dB and $N = 2$. The average MSE of the proposed robust
scheme is less than that of the traditional TMMSE scheme over all
$W$'s. Moreover, compared to the traditional TMMSE, the descending
slope of the proposed robust scheme is flat, which further indicates
that its performance is insensitive to the channel uncertainty.
Especially when $W$ becomes larger, more reliable CSI is available,
therefore closer the two curves get.

Fig. \ref{Fig 4} illustrates the convergence property of the
proposed robust scheme, when $N=2$, $W=100$. The curves of the
average MSE versus the number of iterations needed in different
SNR's are plotted. The higher the SNR is, the more iterations the
proposed scheme runs for to converge. Fortunately, for the most
SNR's, four iterations are big enough to guarantee the convergence.

\section{Conclusion}
In this paper, we investigate a robust linear processing scheme for
the downlink multiuser MIMO system under the consideration of
imperfect CSI. As the traditional downlink multiuser MIMO systems
depend on the instantaneous CSI too much, they suffer poor
performance once the CSI is not accurate enough. In order to deliver
a better performance under the imperfect CSI, an iterative Bayesian
algorithm which explores channel statistics to offer a much more
stable description to the channel state is developed by minimize the
total MSE under the sum power constraint. Numerical simulations
exhibit the proposed robust scheme experiences an obvious
performance gain over the traditional schemes. In addition, the
proposed iterative algorithm has a good convergence property --
after no more than four times of iterations, the algorithm achieves
convergence.

\end{document}